\documentclass[a4paper,11pt]{article}

\usepackage{contribution}

\newcommand{\weblink}[2][]{%
    \ifthenelse{\equal{#1}{}}%
    {\textnormal{\url{#2}}}%
    {\textnormal{\href{#2}{#1}}}%
}

\newcommand{\acknowledgements}[1]{%
  \bigskip\bigskip
  \textsf{\textbf{\Large Acknowledgements}} \\[2ex]
  {#1}
  \bigskip
}

\def\beq{\begin{equation}}
\def\eeq#1{\label{#1}\end{equation}}
\def\eeqn{\end{equation}}

\def\beqa{\begin{eqnarray}}
\def\eeqa#1{\label{#1}\end{eqnarray}}
\def\eeqan{\end{eqnarray}}

\let\bar=\overbar

\def\Dslash{\not{\hbox{\kern-4pt $D$}}}
\def\dslash{\not{\hbox{\kern-2pt $\del$}}}

\def\msb{{\bar{\ssstyle M \kern -1pt S}}}

\newcommand{\contribution}[7][]{%
  \clearpage
  \thispagestyle{plain}
  \ifthenelse{\equal{#1}{}}
  {\hypersetup{pdftitle={#2}}}
  {\hypersetup{pdftitle={#1}}}
  \hypersetup{pdfauthor={{#3} {#4}}}
  {\centering\normalfont\LARGE\bfseries\sffamily #2 \par\nobreak}
  \lhead{}
  \chead{%
    \textit{\footnotesize XIV International Conference on Hadron Spectroscopy
      (\weblink[\textit{hadron2011}]{http://www.hadron2011.de}), 13-17 June 2011, Munich, Germany}%
  }
  \rhead{}
  \bigskip
  \begin{center}
    {#3} {#4}\ifthenelse{\equal{#6}{}}{}{\footnote{\weblink[#6]{mailto:#6}}}
    \ifthenelse{\equal{#7}{}}{}{#7} \\
    \textit{#5}
  \end{center}
  \bigskip
}

\renewcommand{\abstract}[1]{%
  \begin{center}
    \begin{minipage}{0.85\textwidth}
      \begin{footnotesize}
        #1
      \end{footnotesize}
    \end{minipage}
  \end{center}
  \bigskip
}

\begin{document}

{\makeatletter\@ifundefined{c@affiliation}
{\newcounter{affiliation}}{}\makeatother
\newcommand{\affiliation}[2][]{\setcounter{affiliation}{#2}
\ensuremath{{^{\alph{affiliation}}}\text{#1}}}

\contribution[Effective Continuum Thresholds in Sum
Rules]{Unprejudiced Look at Effective Continuum Thresholds in
Borel Dispersive Sum Rules}{Wolfgang}{Lucha}{\affiliation[HEPHY,
Austrian Academy of Sciences, Nikolsdorfergasse 18, A-1050 Vienna,
Austria]{1}\\\affiliation[Faculty of Physics, University of
Vienna, Boltzmanngasse 5, A-1090 Vienna, Austria]{2}\\
\affiliation[SINP, Moscow State University, 119991 Moscow,
Russia]{3}\\\affiliation[INFN, Sezione di Roma Tre, Via della
Vasca Navale 84, I-00146 Roma, Italy]{4}}
{Wolfgang.Lucha@oeaw.ac.at}{\!\!$^,\affiliation{1}$, Dmitri
Melikhov\!\affiliation{1}$^,\affiliation{2}^,\affiliation{3}$, and
Silvano Simula\!\affiliation{4}}

{\em Dispersive sum rules\/} represent long-standing tools for
extracting hadron features from QCD; they are constructed by
evaluating matrix elements of suitable operators (e.g.\
time-ordered products of quark currents) at the level of both
hadron and QCD degrees of freedom. One's ignorance of hadronic
excitations and continuum is circumvented by {\em `quark--hadron
duality'\/}: beyond an {\em `effective threshold'\/} hadron and
QCD contributions are {\em assumed\/} to cancel. We
\cite{LMS:SUE,LMS:FF} estimate the error induced by such
approximation and improve the accuracy of predictions by elevating
our thresholds from constants to functions of momenta and a
parameter $T$ or $\tau$ entering upon {\em Borel transformation\/}
\cite{LMS:ECT}. This move enables us to define {\em dual
correlators\/}, where the QCD member, truncated at effective
threshold $s_{\rm eff},$ {\em exactly\/} counterbalances the
hadronic ground-state member; the form of $s_{\rm eff}$ can be
determined by fitting known hadron~features.

\begin{center}
\includegraphics[width=0.4971\textwidth]{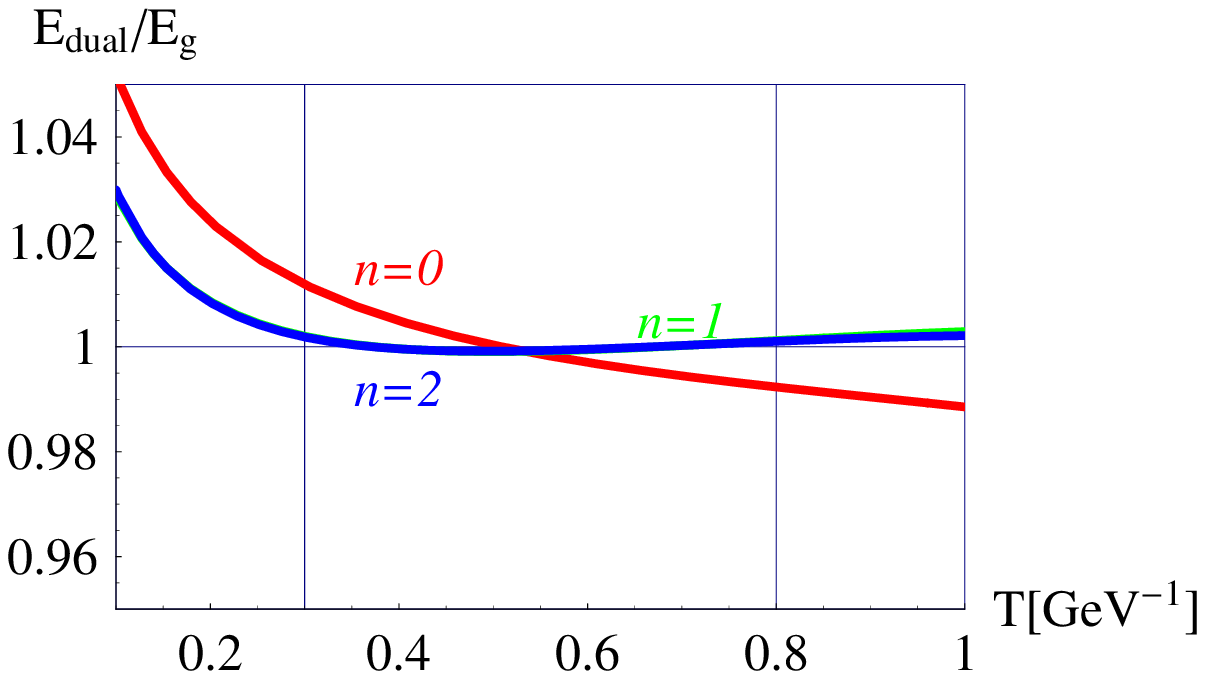}
\includegraphics[width=0.4971\textwidth]{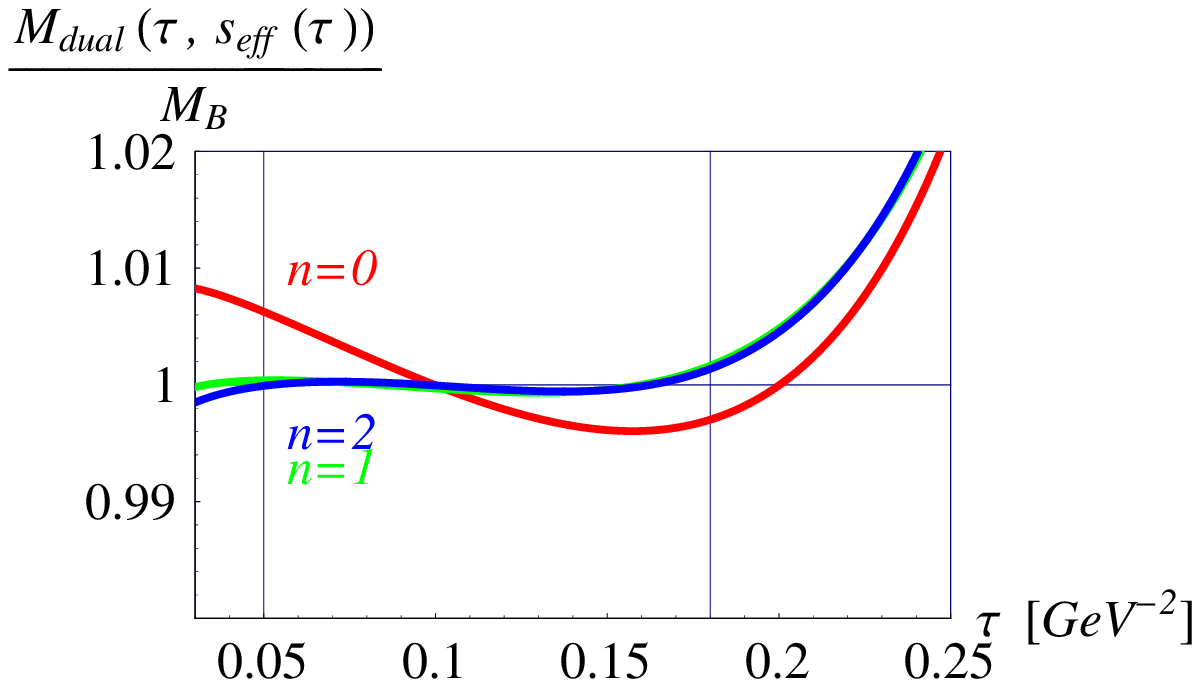}\\
\includegraphics[width=0.4971\textwidth]{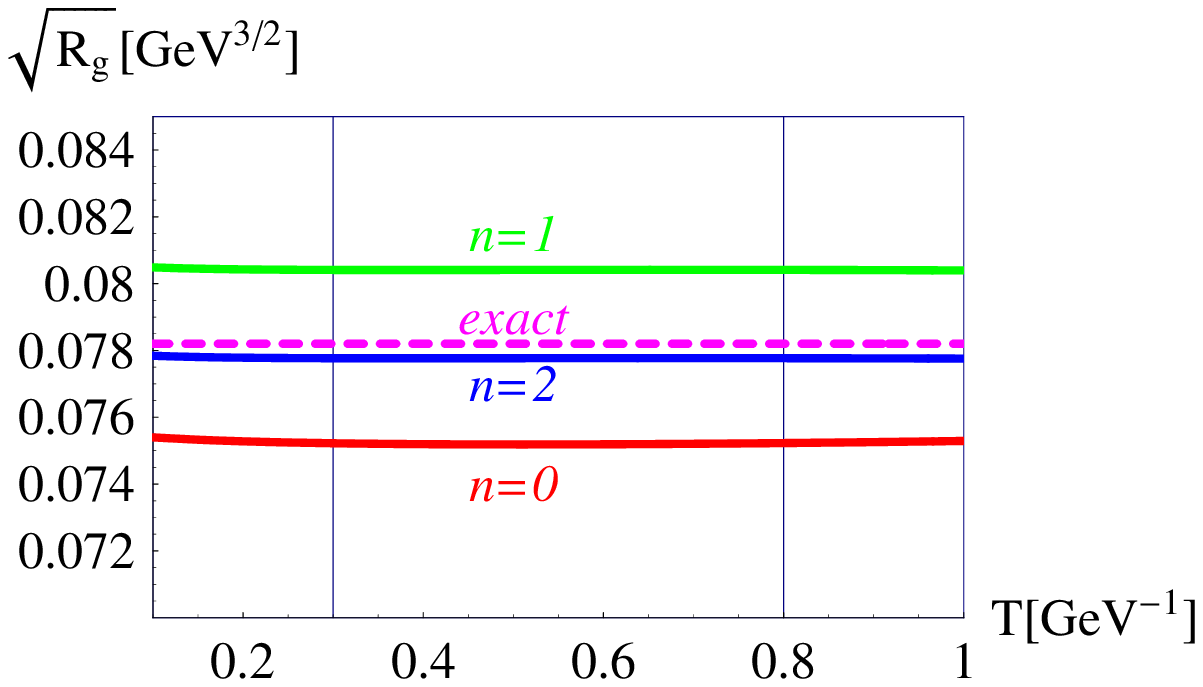}
\includegraphics[width=0.4971\textwidth]{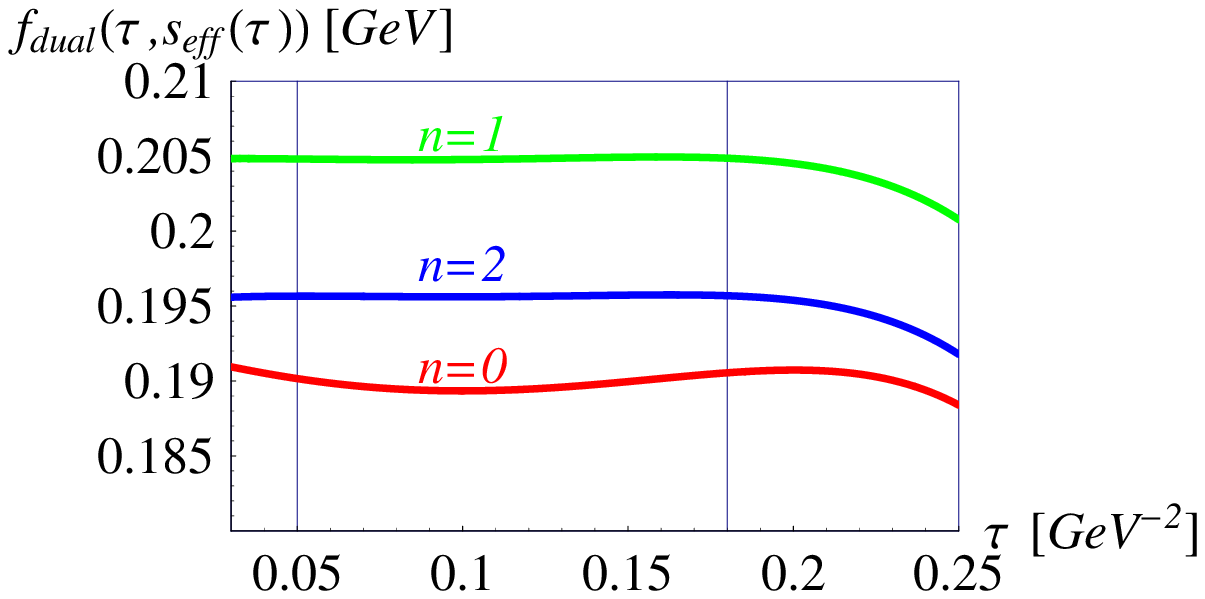}
\end{center}

To scrutinize the applicability of our proposed modified sum-rule
algorithm to QCD \cite{LMS:QCD} we confront extractions of
ground-state decay constants $\sqrt{R_{\rm g}}\equiv|\Psi({\bf
0})|$ in quantum mechanics in terms of related wave functions
$\Psi({\bf x})$ with like extractions of heavy pseudoscalar-meson
decay constants in QCD. The plots depict {\em dual\/} energy
$E_{\rm dual}$ over true $E_{\rm g}$ and decay constant
$\sqrt{R_{\rm g}}$ resulting from the funnel potential describing
heavy-quark bound states \cite{LSG:QBS} (left), and $B$-meson mass
$M_{\rm dual}$ over its experimental value $M_B$ and decay
constant $f_{\rm dual},$ predicted by QCD (right), vs.~the
associated Borel parameter: Adopting polynomial Ans\"atze of
degree $n$ for the effective continuum thresholds, the band
delimited by our $n=1$ and $n=2$ findings will provide an
`educated guess' of the intrinsic errors of bound-state features
such as $\sqrt{R_{\rm g}}.$ The similarity of the procedures in
quantum mechanics and QCD gives us great confidence that our
sum-rule alterations will prove to be successful also in
hadron~phenomenology \cite{LMS:HMDC}.

\vspace{-.8ex}\acknowledgements{D.M.\ acknowledges support by the
Austrian Science Fund (FWF) under Project No.~P22843.}

}

\end{document}